\RequirePackage{fixltx2e}
\documentclass[a4paper]{isp}
\usepackage[american]{babel}
\usepackage{graphicx}
\usepackage{amsmath,amsthm,mathtools}
\usepackage{url}
\usepackage[capitalise,nameinlink]{cleveref}
\usepackage{xcolor}
\def\aap{A\&A\,  }

\def\apj{ApJ\,  }
\def\apjs{ApJS  }
\def\apss{Astrophysics and Space Science  }

\def\snr1993j{SN\,1993J~}
\begin{document}
\pdfgentounicode=1
\title
{
Energy Conservation in the thin  layer approximation:
III. The spherical relativistic case for supernovae 
}
\author{Lorenzo Zaninetti}
\institute{
Physics Department,
 via P. Giuria 1, I-10125 Turin, Italy \\
 \email{zaninetti@ph.unito.it}
}

\maketitle

\begin {abstract}
The theory of the conservation of energy in the
thin layer approximation has been extended  
to  special  relativity.
Four models for the density of the circumstellar medium 
are analyzed, which are represented by constant, power law, exponential
and  Emden (n=5)   profile for density.
The astrophysical results are presented  in a numerical
way, except for a Taylor expansion of the four 
trajectories in the surrounding of the origin.
The free parameters of the models are particularized for 
SN1993j, for which the radius versus time is known.
Some evaluations on the time dilation are presented.
\end{abstract}
{
\bf{Keywords:}
}
supernovae: general,
supernovae: individual (SN1993j),
ISM       : supernova remnants

\section{Introduction}
The production of relativistic electrons 
in the early phase of a supernova (SN) 
is an active field of research.
For example, 
the application
of the non-relativistic perpendicular shocks
to: 
(i)
the  formation of Weibel-type filamentation instabilities,
which  generate magnetic turbulence,
see \cite{Wieland2016};
(ii) 
the shock-surfing acceleration of electrons at the leading edge of
the shock foot and downstream of the shock, see
\cite{Bohdan2019a,Bohdan2019b}; and
(iii) 
to study the magnetic re-connection 
as a  dominant acceleration process for the acceleration of the electrons
, see
\cite{Bohdan2020}.
These approaches use non-relativistic shocks. Therefore,
an  approach in special relativity (SR) is required. 
In this paper, we report some approaches to solve this problem, including:
the relativistic theory of hydrodynamical shocks, 
see \cite{Taub1948}; 
the self-similar spherical solution describing an 
adiabatic ultra relativistic
blast wave, see \cite{Blandford1976}; 
analysis of the reverse shock 
in a dynamical evolution of a relativistic explosion,
see \cite{Yokosawa1984};
evaluation of the  jump conditions in parallel
relativistic collision-less shocks in the absence of Fermi
acceleration,
see \cite{Ellison1991};
the ultra-relativistic shock breakout
with production of  non-thermal emission,
which was investigated by \cite{Nakar2012}; and,
an  analytic description of relativistic 
radiation-mediated shocks with application to SN, 
see \cite{Ohtani2013}.
The astronomical measures  of  
the high-velocity features  in optical spectra of type Ia supernovae
reveals high velocity , $v$,  for the ejecta such as 
$24000\,\frac{km}{s}$, which means $\beta=0.08$; where 
$\beta=\frac{v}{c}$ with $c$ is the light velocity,
see \cite{Zhao2015}.
The measured high velocities in young SNs require
a treatment for the early expansion in the framework 
of SR.
Previous studies analyzed the relativistic conservation 
of momentum  for  the thin layer 
approximation  adopting a 
power law  profile of the density, see \cite{Zaninetti2011a},
and a Lane Emden (n = 5) profile of density,
see \cite{Zaninetti2014f}.
We recall that the relativistic conservation 
of the momentum or the energy 
in the thin
layer approximation is an hypothesis of work
that  should be sustained from the observations,
i.e. the observed trajectory of  
\snr1993j \cite{Marcaide2009}. 
This  paper is structured as follows. In 
Section \ref{secrelativistic}, 
the basic equations of the conservation of the relativistic energy 
for the thin layer approximation are described. 
In Section \ref{secapplications},
the astrophysical results for \snr1993j  
for  four  density profiles  
of the circumstellar medium (CSM) are given.
Finally, time dilation and radioactivity are outlined 
in Section \ref{sectionsparse}.

\section{The relativistic framework}
\label{secrelativistic}

\subsection{Energy conservation} 

The classical conservation of kinetic energy in
spherical coordinates
within the framework of the thin
layer approximation  
when the thermal effects are negligible 
is  
\begin{equation}
\frac{1}{2} M_0(r_0) \,v_0^2 = \frac{1}{2}M(r) \,v^2 
\quad ,
\label{cons_cla_energy}
\end{equation}
where 
$M_0(r_0)$ and $M(r)$ are the swept masses at $r_0$ and $r$, and
$v_0$ and $v$ are the velocities of the thin layer at $r_0$ and $r$;
for further details, see \cite{Zaninetti2020a}.

In SR, the total energy of a particle  is
\begin{equation}
E = M \gamma c^2
\quad ,
\end{equation}
where $M$ is the rest mass,
      $c$ is the light velocity,
      $\gamma$ is the Lorentz factor $\frac{1}{\sqrt{1-\beta^2}}$,
      $\beta=\frac{v}{c}$ and $v$ the velocity.
The relativistic kinetic energy, $E_k$, is
\begin{equation}
E_k = M c^2 (\gamma-1)
\quad ,
\end{equation}
where the rest energy has been subtracted
from the total energy, see formula (23.1) in \cite{Freund2008}.
The relativistic conservation of kinetic energy in 
the thin layer approximation 
in two points ($r_0,v_0$)  and ($r,v$)  is
\begin{equation}
M_0(r_0) c^2 (\gamma_0-1)=M(r) c^2 (\gamma-1)  
\quad ,
\label{cons_rel_energy}
\end{equation}
where $M_0(r_0)$ and $M(r)$ are the swept masses at
$r_0$ and $r$, respectively,
$\gamma_0=\frac{1}{\sqrt{1-\beta_0^2}}$
and   $\beta_0=\frac{v_0}{c}$. 
A Taylor expansion about $v=0$ and $v_0=0$ 
of order three 
for the 
above relativistic conservation law gives the 
classic case given by equation (\ref{cons_cla_energy}).
This fact assures a smooth transition from relativistic
to classical velocities.

\subsection{Constant density}

When the ISM  has a  constant density,
the Lorentz factor  as function of the radius is
\begin{equation}
\beta (r;r_0,\beta_0) =\frac{BN}{BD}
\quad ,
\label{vconstant}
\end{equation}
where 
\begin{eqnarray}
BN=-{r_{{0}}}^{3/2}\Bigg [ {-2\, \left(  \left(  \left( {r}^{3}-\frac{1}{2}\,{r_{{0
}}}^{3} \right) {\beta_{{0}}}^{2}-{r}^{3}+{r_{{0}}}^{3} \right) \sqrt 
{C}+ \left( -{r}^{3}+{r_{{0}}}^{3} \right) {\beta_{{0}}}^{2}+{r}^{3}-{
r_{{0}}}^{3} \right) 
}
\nonumber \\
{
 \left(  \left(  \left( {r}^{6}-2\,{r}^{3}{r_{{0}
}}^{3}+{r_{{0}}}^{6} \right) {\beta_{{0}}}^{2}-{r}^{6}+2\,{r}^{3}{r_{{0
}}}^{3}-2\,{r_{{0}}}^{6} \right) \sqrt {C}+ \left( 2\,{r}^{3}{r_{{0}}}
^{3}-2\,{r_{{0}}}^{6} \right) {\beta_{{0}}}^{2}-2\,{r}^{3}{r_{{0}}}^{3
}+2\,{r_{{0}}}^{6} \right)\Bigg ]^{\frac{1}{2}} }
\, ,
\end{eqnarray}
\begin{eqnarray}
BD=
\sqrt {C}{\beta_{{0}}}^{2}{r}^{6}-2\,\sqrt {C}{\beta_{{0}}}^{2}{r}^{3}
{r_{{0}}}^{3}+\sqrt {C}{\beta_{{0}}}^{2}{r_{{0}}}^{6}+2\,{\beta_{{0}}}
^{2}{r}^{3}{r_{{0}}}^{3}-2\,{\beta_{{0}}}^{2}{r_{{0}}}^{6}
\nonumber \\
-\sqrt {C}{r
}^{6}+2\,\sqrt {C}{r}^{3}{r_{{0}}}^{3}-2\,\sqrt {C}{r_{{0}}}^{6}-2\,{r
}^{3}{r_{{0}}}^{3}+2\,{r_{{0}}}^{6}
\quad,
\end{eqnarray}
and 
\begin{equation}
C=1-{\beta_{{0}}}^{2}
\quad .
\end{equation}
The differential equation that regulates 
the motion can be obtained from the above equation by inserting
$\frac{dr}{dt}=v=\beta\times c$ and $v_0=\beta_0 \times c$,
\begin{equation}
\frac{dr(t;r_0,v_0,c)}{dt}=
\frac{CN}{CD}
\quad ,
\end{equation}
where
\begin{eqnarray}
CN = \nonumber \\
{r_{{0}}}^{3/2}\sqrt {4}\Bigg [ \Big  (    (    ( -\frac{1}{2}\,{c}^{2}+
\frac{1}{2}\,{r_{{0}}}^{2}   )    ( r   ( t   )    ) ^{6}+
   ( {c}^{2}{r_{{0}}}^{3}-{r_{{0}}}^{5}   )    ( r   ( t
   )    ) ^{3}-{r_{{0}}}^{6}   ( {c}^{2}-\frac{1}{2}\,{r_{{0}}}^{2
}   )  \Big  ) \sqrt {{c}^{2}-{r_{{0}}}^{2}}
\nonumber  \\
+   ( -{r_{{0}}}^{
3}{c}^{3}+c{r_{{0}}}^{5}   )    ( r   ( t   ) 
   ) ^{3}+{r_{{0}}}^{6}{c}^{3}-c{r_{{0}}}^{8}   ) \Big   ( 
   \big  (    ( {c}^{2}-{r_{{0}}}^{2}   )    ( r   ( t
   )    ) ^{3}-{r_{{0}}}^{3}   ( {c}^{2}-\frac{1}{2}\,{r_{{0}}}^{2
}   )   \big ) \sqrt {{c}^{2}-{r_{{0}}}^{2}}+
\nonumber \\
   ( -{c}^{3}+c
{r_{{0}}}^{2}   )    ( r   ( t   )    ) ^{3}+{r_{{0}
}}^{3}{c}^{3}-c{r_{{0}}}^{5}  \Big ) \Bigg ] ^{\frac{1}{2}}c
\end{eqnarray}
and 
\begin{eqnarray}
CD =
\sqrt {{c}^{2}-{r_{{0}}}^{2}} \left( r \left( t \right)  \right) ^{6}
{c}^{2}-\sqrt {{c}^{2}-{r_{{0}}}^{2}} \left( r \left( t \right) 
 \right) ^{6}{r_{{0}}}^{2}-2\,\sqrt {{c}^{2}-{r_{{0}}}^{2}} \left( r
 \left( t \right)  \right) ^{3}{c}^{2}{r_{{0}}}^{3}
\nonumber \\
+2\,\sqrt {{c}^{2
}-{r_{{0}}}^{2}} \left( r \left( t \right)  \right) ^{3}{r_{{0}}}^{5}+
2\,\sqrt {{c}^{2}-{r_{{0}}}^{2}}{c}^{2}{r_{{0}}}^{6}-\sqrt {{c}^{2}
-{r_{{0}}}^{2}}{r_{{0}}}^{8}+2\, \left( r \left( t \right)  \right) ^{
3}{c}^{3}{r_{{0}}}^{3}-2\, \left( r \left( t \right)  \right) ^{3}c{
r_{{0}}}^{5}
\nonumber \\
-2\,{r_{{0}}}^{6}{c}^{3}+2\,c{r_{{0}}}^{8}
\quad .
\label{eqndiff_constant}
\end{eqnarray}
This differential does not have an analytical solution
and therefore the solution should be derived in a 
numerical way, except 
about $r=r_0$, where a third-order Taylor series expansion gives 
\begin{equation}
r(t;r_0,v_0,t_0)=
r_{{0}}+v_{{0}} \left( t-t_{{0}} \right) + \frac{3}{2}\,{\frac { \left( c-v_{{0
}} \right)  \left( c+v_{{0}} \right)  \left( {c}^{2}-c\sqrt {{c}^{
2}-{v_{{0}}}^{2}}-{v_{{0}}}^{2} \right)  \left( t-t_{{0}} \right) ^{2}
}{c\sqrt {{c}^{2}-{v_{{0}}}^{2}}r_{{0}}}}
\quad .
\label{rt_taylor_costante}
\end{equation}
\subsection{A power law profile for the density }
\label{section_powerlaw}

The medium 
is supposed to scale as 
\begin{equation}
 \rho (r;r_0)  = \{ \begin{array}{ll}
            \rho_c                      & \mbox {if $r \leq r_0 $ } \\
            \rho_c (\frac{r_0}{r})^{\alpha}    & \mbox {if $r >     r_0 $.}
            \end{array}
\label{piecewisealpha},
\end{equation}
where 
$\rho_c$ is the density at $r=0$,
$r_0$ is the radius after which the density 
starts to decrease
and
$\alpha >0$.

The total mass swept, $ M(r;r_0,\rho_c,\alpha) $,
in the interval [0,r]
is
\begin{eqnarray}
M (r;r_0,\rho_c,\alpha)=
\frac{4}{3}\,\rho_{{c}}\pi\,{r_{{0}}}^{3}-4\,{\frac {{r}^{3}\rho_{{c}}\pi}{
\alpha-3} \left( {\frac {r_{{0}}}{r}} \right) ^{\alpha}}+4\,{\frac {
\rho_{{c}}\pi\,{r_{{0}}}^{3}}{\alpha-3}}
\quad .
\nonumber
\label{massinversepowerlaw}
\end{eqnarray}
The conservation of energy in SR gives the following 
differential equation
\begin{equation}
\frac{dr(t;r_0,v_0,c)}{dt}=
\frac{PN}{PD}
\quad ,
\label{eqndiff_powerlaw} 
\end{equation}
where
\begin{eqnarray}
PN = 
-{r_{{0}}}^{3/2}c
\Bigg [ -54\, \Big  (    ( 1/9\,   (    ( {
\alpha}^{2}-18\,\alpha+72   ) {c}^{2}+   ( {\alpha}^{2}+6\,
\alpha-54   ) {v_{{0}}}^{2}   ) {r_{{0}}}^{6+\alpha}   ( r
   ( t   )    ) ^{-\alpha+3}
\nonumber \\
+2/3\,   ( \alpha-15/2
   ) {r_{{0}}}^{2\,\alpha+3}   ( c+v_{{0}}   )    ( c-v_
{{0}}   )    ( r   ( t   )    ) ^{-2\,\alpha+6}+{r_{
{0}}}^{3\,\alpha}   ( {c}^{2}-{v_{{0}}}^{2}   )    ( r
   ( t   )    ) ^{-3\,\alpha+9}
\nonumber \\
-1/9\,   (    ( {
\alpha}^{2}-18   ) {v_{{0}}}^{2}+   ( \alpha-6   ) ^{2}{c}^
{2}   ) {r_{{0}}}^{9}   )    ( c+v_{{0}}   )    ( c-
v_{{0}}   ) c\sqrt {   ( {c}^{2}-{v_{{0}}}^{2}   ) ^{-1}}-1
/9\,   (    ( {\alpha}^{2}-18\,\alpha+72   ) {c}^{2}
\nonumber \\
-3\,
   ( 6+\alpha   ) {v_{{0}}}^{2}   )    ( c+v_{{0}}
   )    ( c-v_{{0}}   ) {r_{{0}}}^{6+\alpha}   ( r
   ( t   )    ) ^{-\alpha+3}-2/3\,{r_{{0}}}^{2\,\alpha+3}
   (    ( \alpha-15/2   ) {c}^{2}
\nonumber \\ 
+1/4\,   ( \alpha+15
   ) {v_{{0}}}^{2}   )    ( c+v_{{0}}   )    ( c-v_{
{0}}   )    ( r   ( t   )    ) ^{-2\,\alpha+6}-{r_{{0
}}}^{3\,\alpha}   ( c-v_{{0}}   ) ^{2}   ( c+v_{{0}}
   ) ^{2}   ( r   ( t   )    ) ^{-3\,\alpha+9}
\nonumber \\
+1/9\,
{r_{{0}}}^{9}   (    ( \alpha-6   ) ^{2}{c}^{4}-1/6\,
   ( {\alpha}^{3}-3\,{\alpha}^{2}-36\,\alpha+216   ) {v_{{0}}}^
{2}{c}^{2}+3/2\,   ( \alpha+3   ) {v_{{0}}}^{4}   ) 
 \Big  )    ( \alpha-3   )  \Bigg ]^{\frac{1}{2}}
\quad ,
\end{eqnarray}
and
\begin{eqnarray}
PD = 
-6\,\sqrt {{\frac {{c}^{2}}{{c}^{2}-{v_{{0}}}^{2}}}}\alpha\,{c}^{2}{r_
{{0}}}^{6}+6\,\sqrt {{\frac {{c}^{2}}{{c}^{2}-{v_{{0}}}^{2}}}}\alpha\,
{r_{{0}}}^{6}{v_{{0}}}^{2}-{\alpha}^{2}{c}^{2}{r_{{0}}}^{6}+18\,\sqrt 
{{\frac {{c}^{2}}{{c}^{2}-{v_{{0}}}^{2}}}}{c}^{2}{r_{{0}}}^{6}
\nonumber \\
-18\,
\sqrt {{\frac {{c}^{2}}{{c}^{2}-{v_{{0}}}^{2}}}}{r_{{0}}}^{6}{v_{{0}}}
^{2}+6\,\alpha\,{c}^{2}{r_{{0}}}^{6}-18\,{c}^{2}{r_{{0}}}^{6}+9\,{r_{{0
}}}^{6}{v_{{0}}}^{2}+6\,\sqrt {{\frac {{c}^{2}}{{c}^{2}-{v_{{0}}}^{2}}
}}{r_{{0}}}^{\alpha+3} \left( r \left( t \right)  \right) ^{-\alpha+3}
\alpha\,{c}^{2}
\nonumber \\
-6\,\sqrt {{\frac {{c}^{2}}{{c}^{2}-{v_{{0}}}^{2}}}}{r_
{{0}}}^{\alpha+3} \left( r \left( t \right)  \right) ^{-\alpha+3}
\alpha\,{v_{{0}}}^{2}-18\,\sqrt {{\frac {{c}^{2}}{{c}^{2}-{v_{{0}}}^{2
}}}}{r_{{0}}}^{\alpha+3} \left( r \left( t \right)  \right) ^{-\alpha+
3}{c}^{2}
\nonumber  \\
+18\,\sqrt {{\frac {{c}^{2}}{{c}^{2}-{v_{{0}}}^{2}}}}{r_{{0}}
}^{\alpha+3} \left( r \left( t \right)  \right) ^{-\alpha+3}{v_{{0}}}^
{2}+18\,{r_{{0}}}^{\alpha+3} \left( r \left( t \right)  \right) ^{-
\alpha+3}{c}^{2}-18\,{r_{{0}}}^{\alpha+3} \left( r \left( t \right) 
 \right) ^{-\alpha+3}{v_{{0}}}^{2}
\nonumber  \\
-9\,{r_{{0}}}^{2\,\alpha} \left( r
 \left( t \right)  \right) ^{-2\,\alpha+6}{c}^{2}+9\,{r_{{0}}}^{2\,
\alpha} \left( r \left( t \right)  \right) ^{-2\,\alpha+6}{v_{{0}}}^{2
}
\quad .
\end{eqnarray}
A third-order Taylor series expansion gives  
\begin{equation}
r(t;t_0,r_0,v_0)=
r_{{0}}+v_{{0}} \left( t-t_{{0}} \right) -3/2\,{\frac { \left( c-v_{{0
}} \right)  \left( c+v_{{0}} \right)  \left( c-\sqrt {{c}^{2}-{v_{{0}}
}^{2}} \right)  \left( t-t_{{0}} \right) ^{2}}{r_{{0}}c}}
\quad .
\end{equation}

\subsection{An exponential profile}

\label{section_exponential}
We  assume that the medium 
around the SN
scales with the piecewise dependence
\begin{equation}
 \rho (r;r_0)  = \{ \begin{array}{ll}
            \rho_c                        & \mbox {if $r \leq r_0 $ } \\
            \rho_c (\exp{-\frac{r}{b}})    & \mbox {if $r >     r_0 $.}
            \end{array}
\label{piecewiseexp},
\end{equation}
where 
$\rho_c$ is the density at $r=0$
and  
$r_0$ is the radius after which the density 
starts to decrease.
The total mass swept, $ M(r;r_0,\rho_c) $,
in the interval [0,r]
is
\begin{equation}
M (r;r_0,\rho_c,b)=
\frac{4}{3}\,\rho_{{c}}\pi\,{r_{{0}}}^{3}-4\,b \left( 2\,{b}^{2}+2\,br+{r}^{2}
 \right) \rho_{{c}}{{\rm e}^{-{\frac {r}{b}}}}\pi+4\,b \left( 2\,{b}^{
2}+2\,br_{{0}}+{r_{{0}}}^{2} \right) \rho_{{c}}{{\rm e}^{-{\frac {r_{{0
}}}{b}}}}\pi
\quad .
\label{massexponential}
\end{equation}

The conservation of energy in SR gives the following 
differential equation
\begin{equation}
\frac{dr(t;r_0,v_0,c,b)}{dt}=
\frac{EN}{ED}
\quad ,
\label{eqndiffexp}
\end{equation}
where
\begin{eqnarray}
EN = 
{{\rm e}^{\frac{3}{2} {\frac {r   ( t   )
+r_{{0}}}{b}}}}{r_{{0}}}^{\frac{3}{2}}
\Bigg [ {   ( {{\rm e}^{{\frac {r   ( t   ) +r_{{0}}}{b}}}}{
r_{{0}}}^{3}{v_{{0}}}^{2}+12    (    ( -{b}^{2}-br_{{0}}-\frac{1}{2} {
r_{{0}}}^{2}   ) {{\rm e}^{{\frac {r   ( t   ) }{b}}}}+{
{\rm e}^{{\frac {r_{{0}}}{b}}}}   ( {b}^{2}+br   ( t   ) 
}
\nonumber \\
{
+\frac{1}{2}    ( r   ( t   )    ) ^{2}   )    ) b   ( 
v_{{0}}+c   )    ( -v_{{0}}+c   )    ) \sqrt {{c}^{2}-{
v_{{0}}}^{2}}-12    (    ( -{b}^{2}-br_{{0}}
}
\nonumber \\
{
-\frac{1}{2} {r_{{0}}}^{2}
   ) {{\rm e}^{{\frac {r   ( t   ) }{b}}}}+{{\rm e}^{{
\frac {r_{{0}}}{b}}}}   ( {b}^{2}+br   ( t   ) 
+\frac{1}{2} 
   ( r   ( t   )    ) ^{2}   )    ) b   ( v_{
{0}}+c   )    ( -v_{{0}}+c   ) c}
\Bigg ]^{\frac{1}{2}}
\Bigg [  
{   ( {{\rm e}^{{
\frac {2 r   ( t   ) +2 r_{{0}}}{b}}}}{c}^{2}{r_{{0}}}^{6}-72
 {b}^{2}   ( v_{{0}}+c   ) 
}
\nonumber  \\
{
   ( -v_{{0}}+c   )    ( 
   ( {b}^{2}+br_{{0}}+\frac{1}{2} {r_{{0}}}^{2}   )    ( {b}^{2}+br
   ( t   ) +\frac{1}{2}    ( r   ( t   )    ) ^{2}
   ) {{\rm e}^{{\frac {r   ( t   ) +r_{{0}}}{b}}}}-\frac{1}{2} {
{\rm e}^{2 {\frac {r   ( t   ) }{b}}}}   ( {b}^{2}+br_{{0}}
+\frac{1}{2} {r_{{0}}}^{2}   ) ^{2}-\frac{1}{2}    ( {b}^{2}
}
\nonumber  \\
{
+br   ( t
   ) +\frac{1}{2}    ( r   ( t   )    ) ^{2}   ) ^{2}{
{\rm e}^{2 {\frac {r_{{0}}}{b}}}}   )    ) \sqrt {{c}^{2}-{v_
{{0}}}^{2}}-12 b   ( v_{{0}}+c   ) {r_{{0}}}^{3}   ( -v_{{0
}}+c   ) c   (    ( {b}^{2}+br   ( t   ) +\frac{1}{2} 
   ( r   ( t   )    ) ^{2}   ) {{\rm e}^{{\frac {r
   ( t   ) +2 r_{{0}}}{b}}}}
}
\nonumber  \\
{
-{{\rm e}^{{\frac {2 r   ( t
   ) +r_{{0}}}{b}}}}   ( {b}^{2}+br_{{0}}+\frac{1}{2} {r_{{0}}}^{2}
   )    ) }
\Bigg ]^{\frac{1}{2}}
\quad,
\end{eqnarray}
and
\begin{eqnarray}
ED = 
{\frac {1}{c} \Bigg  [ \Big   ( -72\,   ( -v_{{0}}+c   )    ( {
b}^{2}+br_{{0}}+\frac{1}{2}\,{r_{{0}}}^{2}   )    ( v_{{0}}+c   ) {
b}^{2}   ( {b}^{2}+br   ( t   ) +\frac{1}{2}\,   ( r   ( t
   )    ) ^{2}   ) {{\rm e}^{{\frac {2\,r_{{0}}+2\,r
   ( t   ) }{b}}}}
}
\nonumber \\
{
+{{\rm e}^{{\frac {3\,r_{{0}}+3\,r   ( t
   ) }{b}}}}{c}^{2}{r_{{0}}}^{6}+36\,   ( -v_{{0}}+c   ) 
   ( {{\rm e}^{{\frac {3\,r_{{0}}+r   ( t   ) }{b}}}}
   ( {b}^{2}+br   ( t   ) +\frac{1}{2}\,   ( r   ( t   ) 
   ) ^{2}   ) ^{2}
}
\nonumber \\
{
+   ( {b}^{2}+br_{{0}}+\frac{1}{2}\,{r_{{0}}}^{2}
   ) ^{2}{{\rm e}^{{\frac {3\,r   ( t   ) +r_{{0}}}{b}}}}
   )    ( v_{{0}}+c   ) {b}^{2}   ) \sqrt {{c}^{2}-{v_{
{0}}}^{2}}-12\,{r_{{0}}}^{3}   ( -v_{{0}}+c   )    ( v_{{0}}
+c   ) 
} \times
\nonumber \\
\times
{
b   (    ( -{b}^{2}-br_{{0}}-\frac{1}{2}\,{r_{{0}}}^{2}
   ) {{\rm e}^{{\frac {2\,r_{{0}}+3\,r   ( t   ) }{b}}}}+{
{\rm e}^{{\frac {3\,r_{{0}}+2\,r   ( t   ) }{b}}}}   ( {b}^{
2}+br   ( t   ) +\frac{1}{2}\,   ( r   ( t   )    ) ^{2}
   )  \Big  ) c   \Bigg ] }
\quad  .
\end{eqnarray}

A third-order Taylor series expansion gives  
\begin{eqnarray}
r(t;t_0,r_0,v_0,b)=
r_{{0}}+v_{{0}}   ( t-t_{{0}}   ) +3/2\,{\frac {   ( -v_{{0}
}+c   )    ( v_{{0}}+c   )    ( {c}^{2}-c\sqrt {{c}^{2}-
{v_{{0}}}^{2}}-{v_{{0}}}^{2}   )    ( t-t_{{0}}   ) ^{2}}{c
\sqrt {{c}^{2}-{v_{{0}}}^{2}}r_{{0}}}{{\rm e}^{-{\frac {r_{{0}}}{b}}}}
}
\nonumber \\
+{\frac {v_{{0}}   ( v_{{0}}+c   )    ( -v_{{0}}+c   ) 
   ( t-t_{{0}}   ) ^{3}}{b{r_{{0}}}^{2}{c}^{2}} \Big   (    ( 
\sqrt {{c}^{2}-{v_{{0}}}^{2}}-c   )    ( b-\frac{1}{2}\,r_{{0}}
   ) c{{\rm e}^{{\frac {r_{{0}}}{b}}}}
}
\nonumber \\
{
+6\,b   ( c\sqrt {{c}^{2}
-{v_{{0}}}^{2}}-{c}^{2}+3/4\,{v_{{0}}}^{2}   )  \Big  ) {\frac {1}
{\sqrt {{{\rm e}^{4\,{\frac {r_{{0}}}{b}}}}}}}}
\quad .
\label{rt_taylor_exp}
\end{eqnarray}

\subsection{Emden profile}

\label{section_emden}
We  assume that the medium 
around the SN
scales as a $n=5$ Emden profile, 
\cite{Lane1870,Emden1907,Zaninetti2014f},
\begin{equation}
 \rho (r;r_0)  = \{ \begin{array}{ll}
            \rho_c                        & \mbox {if $r \leq r_0 $ } \\
            \rho_c 
\frac{1}
{\left( 1+\frac{1}{3}\,{\frac {{r}^{2}}{{b}^{2}}} \right) ^{5/2}}   & \mbox {if $r >     r_0 $.}
            \end{array}
\label{piecewiseemden},
\end{equation}
where 
$\rho_c$ is the density at $r=0$
and  $b$ is the scale.

The total mass swept, $ M(r;r_0,\rho_c) $,
in the interval [0,r]
is
\begin{equation}
M (r;r_0,\rho_c,b)=
\frac{4}{3}\,\rho_{{c}}\pi\,{r_{{0}}}^{3}+4\,{\frac {{b}^{3}{r}^{3}\rho_{{c}}
\sqrt {3}\pi}{ \left( 3\,{b}^{2}+{r}^{2} \right) ^{3/2}}}-4\,{\frac {{
b}^{3}{r_{{0}}}^{3}\rho_{{c}}\sqrt {3}\pi}{ \left( 3\,{b}^{2}+{r_{{0}}
}^{2} \right) ^{3/2}}}
\quad  .
\end{equation}

The conservation of energy in SR gives the following 
differential equation
\begin{equation}
\frac{dr(t;r_0,v_0,c,b)}{dt}=
\frac{DN}{DD}
\quad ,
\label{eqndiffemden}
\end{equation}
where
\begin{eqnarray}
DN = 
243\,c{r_{{0}}}^{3/2} \Bigg [   ( {b}^{2}+\frac{1}{3}\,   ( r   ( t
   )    ) ^{2}   )    ( \frac{1}{3}\,   ( v_{{0}}+c   ) 
{r_{{0}}}^{3}   (    ( r   ( t   )    ) ^{3}   ( {
b}^{2}+\frac{1}{3}\,{r_{{0}}}^{2}   ) ^{2}\sqrt {3\,{b}^{2}+   ( r
   ( t   )    ) ^{2}}
\nonumber \\
-{r_{{0}}}^{3}   ( {b}^{2}+\frac{1}{3}\,
   ( r   ( t   )    ) ^{2}   ) ^{2}\sqrt {3\,{b}^{2}
+{r_{{0}}}^{2}}   )    ( {b}^{2}+\frac{1}{3}\,   ( r   ( t
   )    ) ^{2}   ) {b}^{3}   ( {b}^{2}+\frac{1}{3}\,{r_{{0}}}^{
2}   ) c   ( -v_{{0}}+c   ) \sqrt {3}\sqrt {   ( {c}^{2}
-{v_{{0}}}^{2}   ) ^{-1}}
\nonumber \\
-\frac{1}{3}\,   ( r   ( t   ) 
   ) ^{3}   ( v_{{0}}+c   ) {r_{{0}}}^{3}   ( {b}^{2}+1/
3\,   ( r   ( t   )    ) ^{2}   ) {b}^{6}   ( {b}
^{2}+\frac{1}{3}\,{r_{{0}}}^{2}   )    ( -v_{{0}}+c   ) \sqrt {3\,{
b}^{2}+{r_{{0}}}^{2}}\sqrt {3\,{b}^{2}+   ( r   ( t   ) 
   ) ^{2}}
\nonumber \\
+   (    ( \frac{1}{2}\,{c}^{2}-\frac{1}{2}\,{v_{{0}}}^{2}
   ) {b}^{12}+{r_{{0}}}^{2}   ( \frac{1}{2}\,{c}^{2}-\frac{1}{2}\,{v_{{0}}}^{2}
   ) {b}^{10}+{r_{{0}}}^{4}   ( 1/6\,{c}^{2}-1/6\,{v_{{0}}}^{2}
   ) {b}^{8}
\nonumber \\
+1/18\,   ( {c}^{2}-2/3\,{v_{{0}}}^{2}   ) {r_{
{0}}}^{6}{b}^{6}+{\frac {{b}^{4}{c}^{2}{r_{{0}}}^{8}}{54}}+{\frac {{b}
^{2}{c}^{2}{r_{{0}}}^{10}}{162}}+{\frac {{c}^{2}{r_{{0}}}^{12}}{1458}}
   )    ( r   ( t   )    ) ^{6}+\frac{1}{3}\,   ( 
   ( {c}^{2}-\frac{1}{2}\,{v_{{0}}}^{2}   ) {b}^{6}
\nonumber \\
+\frac{1}{2}\,{b}^{4}{c}^{2}
{r_{{0}}}^{2}+1/6\,{b}^{2}{c}^{2}{r_{{0}}}^{4}+{\frac {{c}^{2}{r_{{0}}
}^{6}}{54}}   ) {r_{{0}}}^{6}{b}^{2}   ( r   ( t   ) 
   ) ^{4}+   (    ( {c}^{2}-\frac{1}{2}\,{v_{{0}}}^{2}   ) {b}^{
6}+\frac{1}{2}\,{b}^{4}{c}^{2}{r_{{0}}}^{2}+1/6\,{b}^{2}{c}^{2}{r_{{0}}}^{4}
\nonumber \\
+{
\frac {{c}^{2}{r_{{0}}}^{6}}{54}}   ) {r_{{0}}}^{6}{b}^{4}   ( 
r   ( t   )    ) ^{2}+   (    ( {c}^{2}-\frac{1}{2}\,{v_{{0}}
}^{2}   ) {b}^{6}+\frac{1}{2}\,{b}^{4}{c}^{2}{r_{{0}}}^{2}+1/6\,{b}^{2}{c}
^{2}{r_{{0}}}^{4}+{\frac {{c}^{2}{r_{{0}}}^{6}}{54}}   ) {r_{{0}}}
^{6}{b}^{6}   )\times    
\nonumber \\
( 2/3\,   ( v_{{0}}+c   )    ( 
   ( r   ( t   )    ) ^{3}   ( {b}^{2}+\frac{1}{3}\,{r_{{0}}}
^{2}   ) ^{2}\sqrt {3\,{b}^{2}+   ( r   ( t   ) 
   ) ^{2}}
\nonumber \\
-{r_{{0}}}^{3}   ( {b}^{2}+\frac{1}{3}\,   ( r   ( t
   )    ) ^{2}   ) ^{2}\sqrt {3\,{b}^{2}+{r_{{0}}}^{2}}
   ) {b}^{3}c   ( -v_{{0}}+c   ) \sqrt {3}\sqrt {   ( {c
}^{2}-{v_{{0}}}^{2}   ) ^{-1}}
\nonumber \\
-2/3\,   ( r   ( t   ) 
   ) ^{3}   ( v_{{0}}+c   ) {b}^{3}   ( {b}^{2}+\frac{1}{3}\,{r_
{{0}}}^{2}   ) ^{2}   ( -v_{{0}}+c   ) \sqrt {9\,{b}^{2}+3
\,   ( r   ( t   )    ) ^{2}}
+{r_{{0}}}^{3}   ( {b}^{
2}+\frac{1}{3}\,   ( r   ( t   )    ) ^{2}   ) ^{2}   ( 2
/3\,{b}^{3}\times
\nonumber \\
( -v_{{0}}+c   )    ( v_{{0}}+c   ) \sqrt 
{9\,{b}^{2}+3\,{r_{{0}}}^{2}}+   ( {b}^{2}+\frac{1}{3}\,{r_{{0}}}^{2}
   ) ^{2}{v_{{0}}}^{2}   )    ) \Bigg ]^{\frac{1}{2}}\sqrt {18\,{b}^{2}+6\,{r_
{{0}}}^{2}}
\quad ,
\end{eqnarray}
and
\begin{eqnarray}
DN = 
-486\,   ( v_{{0}}+c   ) {r_{{0}}}^{3}  \Bigg [ (    ( r   ( 
t   )    ) ^{3}   ( {b}^{2}+\frac{1}{3}\,{r_{{0}}}^{2}   ) ^{2}
\sqrt {3\,{b}^{2}+   ( r   ( t   )    ) ^{2}}
\nonumber\\
-{r_{{0}}}^
{3}   ( {b}^{2}+\frac{1}{3}\,   ( r   ( t   )    ) ^{2}
   ) ^{2}\sqrt {3\,{b}^{2}+{r_{{0}}}^{2}}   )    ( {b}^{2}+
\frac{1}{3}\,   ( r   ( t   )    ) ^{2}   ) {b}^{3}   ( {
b}^{2}+\frac{1}{3}\,{r_{{0}}}^{2}   ) c   ( -v_{{0}}+c   ) \sqrt {3
}\sqrt {   ( {c}^{2}-{v_{{0}}}^{2}   ) ^{-1}}
\nonumber\\
+486\,   ( r
   ( t   )    ) ^{3}   ( v_{{0}}+c   ) {r_{{0}}}^{3}
   ( {b}^{2}+\frac{1}{3}\,   ( r   ( t   )    ) ^{2}   ) 
{b}^{6}   ( {b}^{2}+\frac{1}{3}\,{r_{{0}}}^{2}   )    ( -v_{{0}}+c
   ) \sqrt {3\,{b}^{2}+{r_{{0}}}^{2}}\sqrt {3\,{b}^{2}+   ( r
   ( t   )    ) ^{2}}
\nonumber \\
+   (    ( -729\,{c}^{2}+729\,{v
_{{0}}}^{2}   ) {b}^{12}
\nonumber\\
+   ( -729\,{c}^{2}+729\,{v_{{0}}}^{2}
   ) {r_{{0}}}^{2}{b}^{10}+   ( -243\,{c}^{2}+243\,{v_{{0}}}^{2
}   ) {r_{{0}}}^{4}{b}^{8}+   ( -81\,{c}^{2}+54\,{v_{{0}}}^{2}
   ) {r_{{0}}}^{6}{b}^{6}-27\,{b}^{4}{c}^{2}{r_{{0}}}^{8}
\nonumber\\
-9\,{b}^{
2}{c}^{2}{r_{{0}}}^{10}-{c}^{2}{r_{{0}}}^{12}   )    ( r
   ( t   )    ) ^{6}+   (    ( -486\,{c}^{2}+243\,{v_
{{0}}}^{2}   ) {r_{{0}}}^{6}{b}^{8}-243\,{b}^{6}{c}^{2}{r_{{0}}}^{
8}-81\,{b}^{4}{c}^{2}{r_{{0}}}^{10}
\nonumber\\
-9\,{b}^{2}{c}^{2}{r_{{0}}}^{12}
   )    ( r   ( t   )    ) ^{4}-1458\,   ( 
   ( {c}^{2}-\frac{1}{2}\,{v_{{0}}}^{2}   ) {b}^{6}+\frac{1}{2}\,{b}^{4}{c}^{2}
{r_{{0}}}^{2}+1/6\,{b}^{2}{c}^{2}{r_{{0}}}^{4}+{\frac {{c}^{2}{r_{{0}}
}^{6}}{54}}   ) {r_{{0}}}^{6}{b}^{4}   ( r   ( t   ) 
   ) ^{2}
\nonumber\\
-1458\,   (    ( {c}^{2}-\frac{1}{2}\,{v_{{0}}}^{2}
   ) {b}^{6}+\frac{1}{2}\,{b}^{4}{c}^{2}{r_{{0}}}^{2}+1/6\,{b}^{2}{c}^{2}{
r_{{0}}}^{4}+{\frac {{c}^{2}{r_{{0}}}^{6}}{54}} \Bigg  ] {r_{{0}}}^{6}{
b}^{6}
\quad .
\end{eqnarray}
A third-order Taylor series expansion gives  
\begin{eqnarray}
r(t;t_0,r_0,v_0,b)=
r_{{0}}+v_{{0}} \left( t-t_{{0}} \right) 
\nonumber \\
-{\frac { \left( -27\,v_{{0}}
+27\,c \right)  \left( v_{{0}}+c \right)  \left( c\sqrt {{c}^{2}-{v_{{0
}}}^{2}}-{c}^{2}+{v_{{0}}}^{2} \right) \sqrt {3}{b}^{5} \left( t-t_{{0
}} \right) ^{2}}{2\, \left( 3\,{b}^{2}+{r_{{0}}}^{2} \right) ^{5/2}
\sqrt {{c}^{2}-{v_{{0}}}^{2}}r_{{0}}c}}
\quad .
\label{rt_taylor_emdequation}
\end{eqnarray}

\section{Astrophysical observations}

\label{secapplications}
We now analyze in detail
the case of  \snr1993j; note that 
the radius in  pc and the elapsed time  in years
can be found in  Table 1 of  \cite{Marcaide2009}.

\subsection{Statistics}

\begin{figure*}
\begin{center}
\includegraphics[width=5cm,angle=-90]{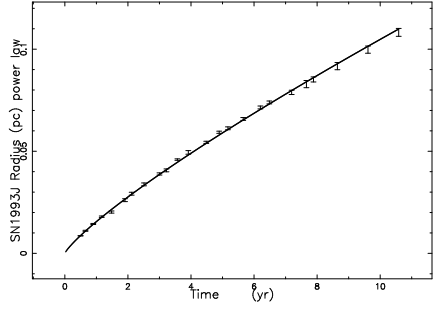}
\end {center}
\caption
{
Radius as fitted by a power law 
(full line)
and astronomical data of \snr1993j with
vertical error bars.
The parameters of the fit are  $r_p=0.0155\,pc$ 
and $\alpha_p=0.828$,  which gives $\chi^2=6387$.
}
\label{powerlaw}
    \end{figure*}

A  test for 
the quality of the fits is represented by the
merit function
$\chi^2$
\begin{equation}
\chi^2  =
\sum_j \frac {(r_{th} -r_{obs})^2}
             {\sigma_{obs}^2}
\quad ,
\nonumber
\label{chisquare}
\end{equation}
where  $r_{th}$, $r_{obs}$ and $\sigma_{obs}$
are the theoretical radius, the observed radius  and
the observed uncertainty, respectively.
A   fit
can be done by assuming  a  power law
dependence  of the type
\begin{equation}
r(t) = r_p t^{\alpha_p}
\nonumber
\label{rpower}
\quad ,
\end{equation}
where the  two parameters $r_p$ and  $\alpha_p$,
as well as their  uncertainties
can be found
using the recipes  suggested in
\cite{Zaninetti2011a}.
Figure \ref{powerlaw} reports the power law fit to the data.

\subsection{Theoretical fits}

\begin{figure*}
\begin{center}
\includegraphics[width=5cm,angle=-90]{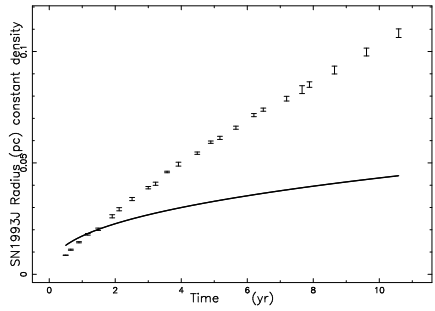}
\end {center}
\caption
{
Theoretical radius  in the case of constant density 
(full line)
and astronomical data of \snr1993j with
vertical error bars.
The parameters of the model  are
$r_0=10^{-3}\,pc$,
$t_0=3.6\,10^{-3}\,yr$,
$\beta_0 = 0.9$
which gives $\chi^2=28208$.
}
\label{rt_constant}
    \end{figure*}

\begin{figure*}
\begin{center}
\includegraphics[width=5cm,angle=-90]{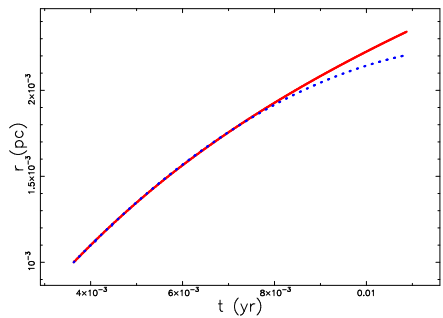}
\end {center}
\caption
{
Numerical solution (full-red line) and  
Taylor solution 
(blue-dashed  line)
for the constant density, 
parameters  as  Figure \ref{rt_constant}.
}
\label{taylor_costante}
    \end{figure*}

In the case of a {\it constant}   profile of
density, we present  a numerical solution   as given
by the numerical integration 
of the differential  equation (\ref{eqndiff_constant}).
Figure \ref{rt_constant} displays 
the theoretical model versus the astronomical data.
Figure \ref{taylor_costante} presents
the Taylor approximation of the trajectory 
as  given by (\ref{rt_taylor_costante})
in the
restricted range of time  $[0.001\,yr-0.02\,yr]$.

\begin{figure*}
\begin{center}
\includegraphics[width=5cm,angle=-90]{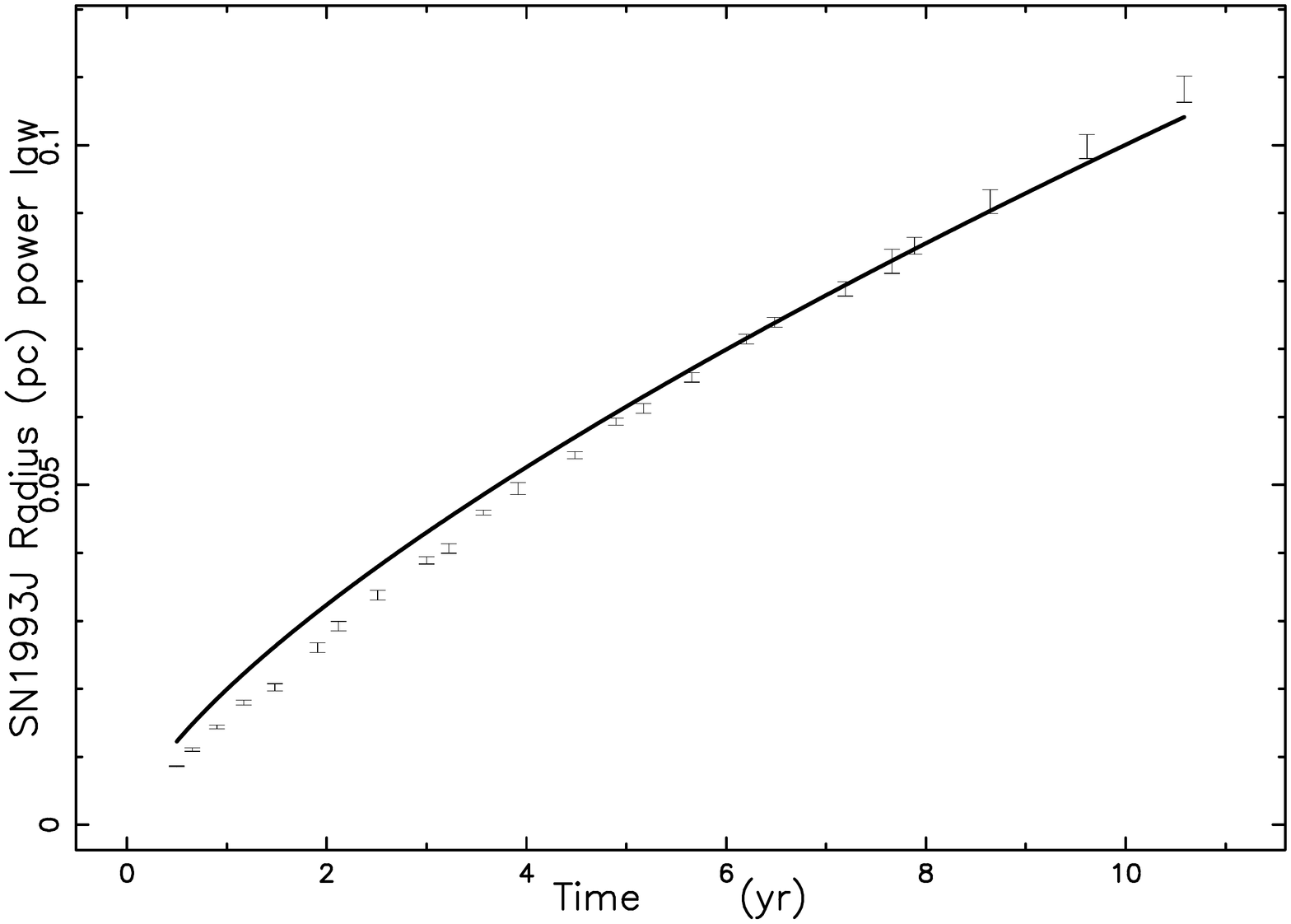}
\end {center}
\caption
{
Theoretical radius  in the case of a power law profile 
(full line)
and astronomical data of \snr1993j with
vertical error bars.
The parameters of the model  are
$r_0=2\,10^{-5}\,pc$,
$t_0=7.2\,10{-5}\,yr$,
$\beta_0 = 0.9$,
$\alpha$=2.15, 
which gives $\chi^2=3777$.
}
\label{rt_power}
    \end{figure*}

In the case of a {\it power law}   profile for   
density, we present  a numerical solution   as given
by the numerical integration 
of the differential  equation (\ref{eqndiff_powerlaw} ).
Figure \ref{rt_power} displays 
the theoretical model.

\begin{figure*}
\begin{center}
\includegraphics[width=5cm,angle=-90]{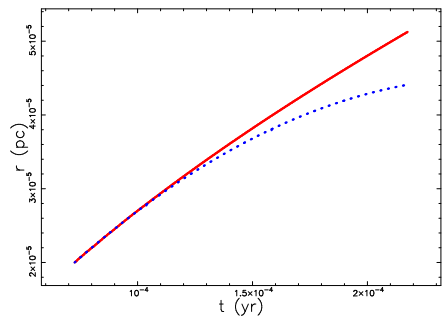}
\end {center}
\caption
{
Numerical solution (full-red line) and  
Taylor solution 
(blue-dashed  line)
for the power law profile, 
the parameters are the same  as in  Figure \ref{rt_power}.
}
\label{rt_power_taylor}
    \end{figure*}

Figure \ref{rt_power_taylor} presents
the Taylor approximation of the trajectory 
as  given by (\ref{rt_power_taylor})
in the
restricted range of time  $[7.2\,10^{-5}\,yr-2.2\,10^{-4}\,yr]$.

\begin{figure*}
\begin{center}
\includegraphics[width=5cm,angle=-90]{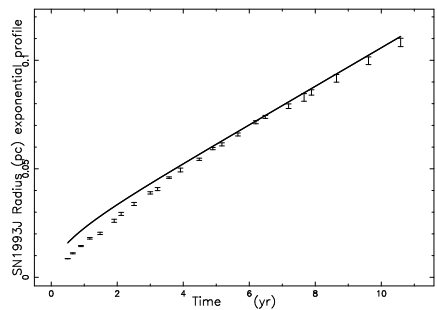}
\end {center}
\caption
{
Theoretical radius  in the case of an exponential profile 
(full line)
and astronomical data of \snr1993j with
vertical error bars.
The parameters of the model  are
$r_0=10^{-3}\,pc$,
$t_0=3.6\,10{-3}\,yr$,
$\beta_0 = 0.9$ and 
$b=8\,10^{-3}\,pc$
which gives $\chi^2=13145$.
}
\label{rt_exp}
    \end{figure*}

In the case of an {\it exponential}    profile for   
density, we present  a numerical solution   as given
by the numerical integration 
of the differential  equation 
(\ref{eqndiffexp} );
Figure \ref{rt_exp} displays 
the theoretical model.   

\begin{figure*}
\begin{center}
\includegraphics[width=5cm,angle=-90]{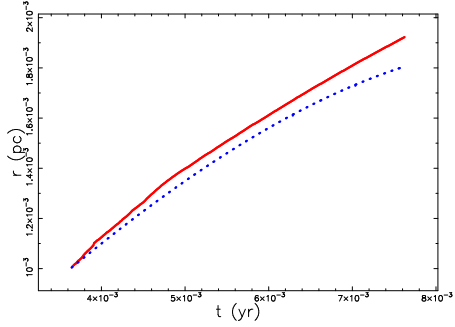}
\end {center}
\caption
{
Numerical solution (full-red line) and  
Taylor solution 
(blue-dashed  line)
for the exponential profile, 
parameters  as  Figure \ref{rt_exp}.
}
\label{exp_taylor_rt}
    \end{figure*}

Figure \ref{exp_taylor_rt} presents
the Taylor approximation of the trajectory 
as  given by (\ref{rt_taylor_exp})
in the
restricted range of time  $[10^{-3}\,yr-8\,10^{-3}\,yr]$.

\begin{figure*}
\begin{center}
\includegraphics[width=5cm,angle=-90]{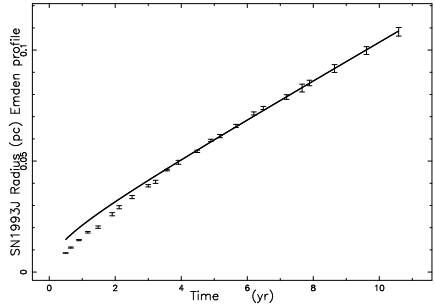}
\end {center}
\caption
{
Theoretical radius  in the case of an Emden profile 
(full line)
and astronomical data of \snr1993j with
vertical error bars.
The parameters of the model  are
$r_0=10^{-3}\,pc$,
$t_0=3.6\,10{-3}\,yr$,
$\beta_0 = 0.9$ and 
$b=8.6\,10^{-3}\,pc$
which gives $\chi^2=8888$.
}
\label{rt_emd}
    \end{figure*}

In the case of an {\it Emden}    profile for   
density, we present  a numerical solution   as given
by the numerical integration 
of the differential  equation 
(\ref{eqndiffemden} ).
Figure \ref{rt_emd} displays 
the theoretical model.   

\begin{figure*}
\begin{center}
\includegraphics[width=5cm,angle=-90]{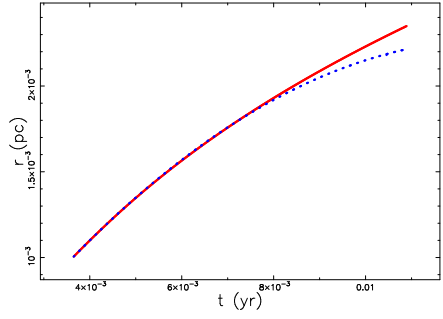}
\end {center}
\caption
{
Numerical solution (full-red line) and  
Taylor solution 
(blue-dashed  line)
for the Emden profile, the
parameters are the same as in Figure \ref{rt_emd}.
}
\label{rt_taylor_emd}
    \end{figure*}

Figure \ref{rt_taylor_emd} presents
the Taylor approximation of the trajectory 
as  given by (\ref{rt_taylor_emdequation})
in the
restricted range of time  $[10^{-3}\,yr-2\,10^{-2}\,yr]$.

\section{Sparse effects}

\label{sectionsparse}
\subsection{Time dilation}

\begin{figure*}
\begin{center}
\includegraphics[width=5cm]{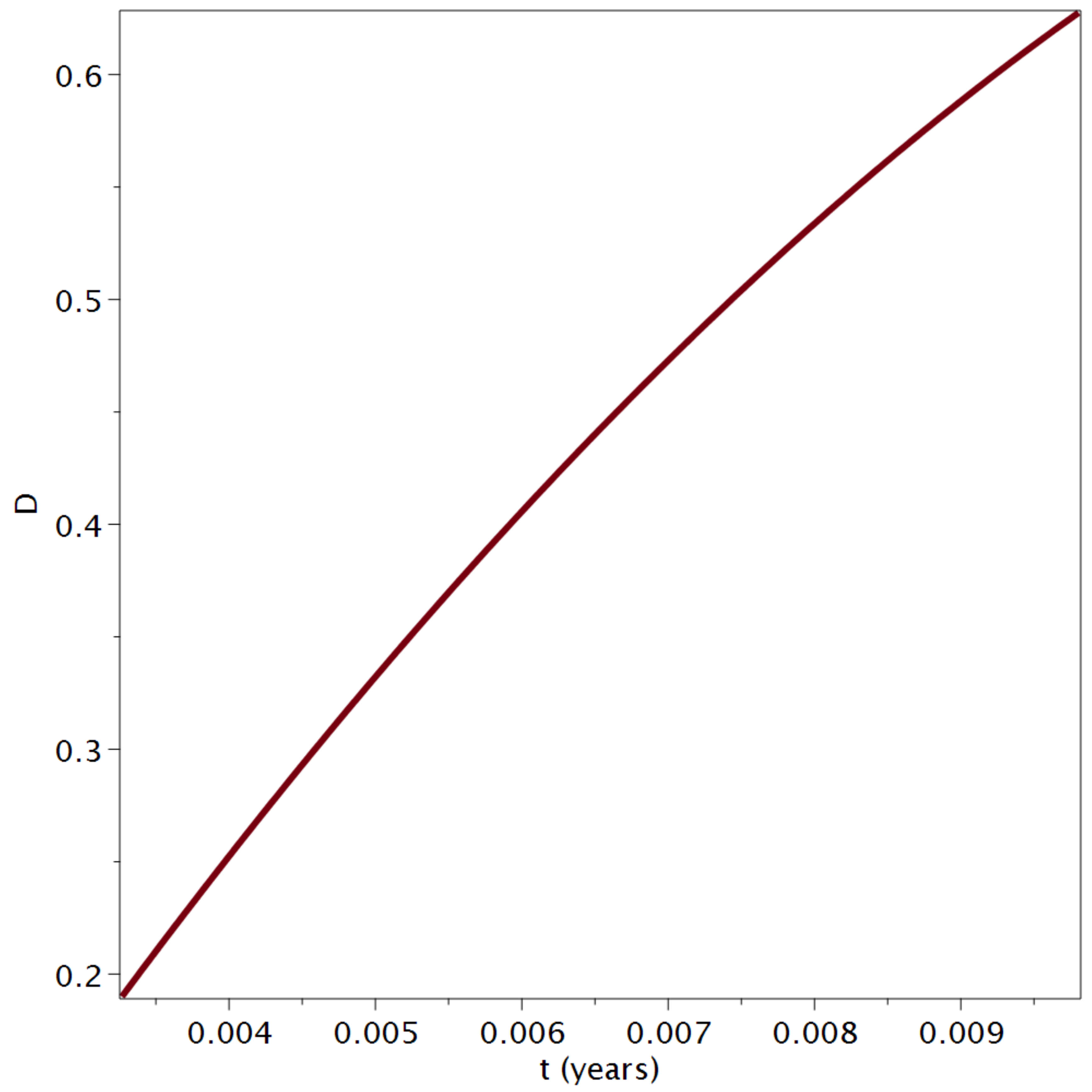}
\end {center}
\caption
{
Time dilation  as represented by $D$
as a function of time (in years) 
for the Emden profile, the 
parameters are the same as in Figure \ref{rt_emd}.
}
\label{dgraph}
    \end{figure*}

\begin{figure*}
\begin{center}
\includegraphics[width=5cm]{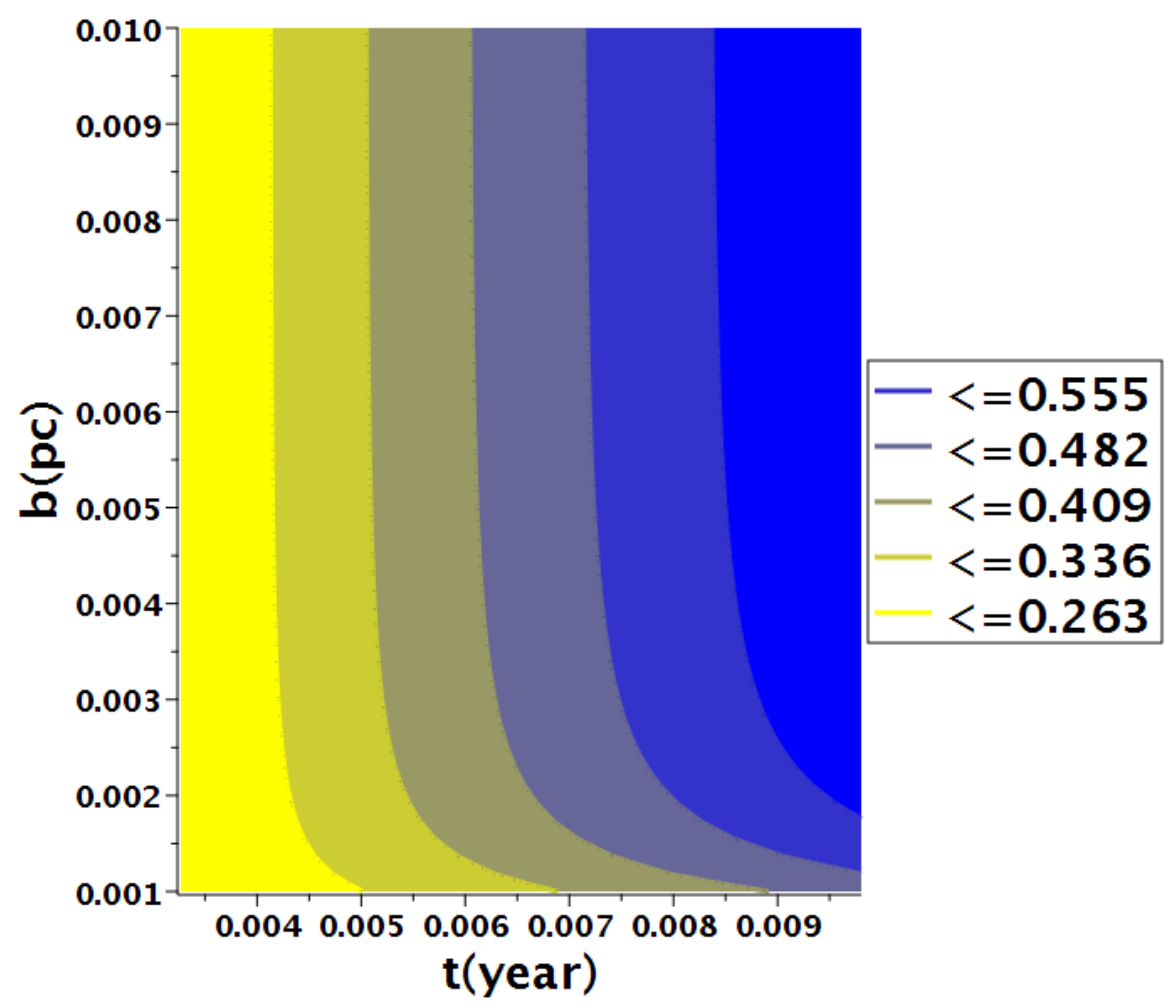}
\end {center}
\caption
{
Map of time dilation, $D$,
as a function of time (in years)
and the scaling $b$  
for the Emden profile, the 
parameters  are the same as in Figure \ref{rt_emd}.
}
\label{matrixgraph}
    \end{figure*}

For an   observer who moves on the expanding shell,
the proper time $\tau^*$ is
\begin{equation}
\tau^* = \int_{t_0}^{t} \frac {dt}{\gamma} =  \int_{t_0}^{t}
\sqrt{1-\beta^2} dt
\quad ,
\nonumber
\end{equation}
see  \cite{Einstein1905}.
Let us take the example of an Emden profile  
with the initial trajectory characterized  by
the Taylor expansion given by equation (\ref{rt_taylor_emdequation}).
The value of $\beta$ as given by the Taylor expansion is
\begin{equation}
\beta(t;r_0,\beta_0,t_0,b)=
\frac{1}{c}
\frac
{d}
{dt} r(t;t_0,r_0,v_0,b)
=
\frac
{BEN}
{
 \left( 3  {b}^{2}+{r_{{0}}}^{2} \right) ^{5/2}\sqrt {-{\beta_{{0}}}^{
2}+1}r_{{0}}
}
\quad , 
\end{equation}
where 
\begin{eqnarray}
BEN=
9\,\sqrt {-{\beta_{{0}}}^{2}+1}r_{{0}} \left( {b}^{2}+1/3\,{r_{{0}}}^{
2} \right) ^{2}\beta_{{0}}\sqrt {3\,{b}^{2}+{r_{{0}}}^{2}}+27\,{b}^{5}
c\sqrt {3} \left( \beta_{{0}}-1 \right)  \left( \beta_{{0}}+1 \right) 
\nonumber \\
 \left( {\beta_{{0}}}^{2}+\sqrt {-{\beta_{{0}}}^{2}+1}-1 \right) 
 \left( t-{\it t0} \right) 
\quad .
\end{eqnarray}
The time dilation can be evaluated once the 
following integral is 
done 
\begin{equation}
F(t;r_0,\beta_0,t_0,b) 
= \int \sqrt{1-\beta(t;r_0,\beta_0,t_0,b)^2} dt
\quad ,
\nonumber
\end{equation}
which is
\begin{equation}
F(t;r_0,\beta_0,t_0,b)   =
{\frac {{{\it FA}}^{3}\sqrt {3}}{{\it FB}}}+t
\quad ,
\end{equation}
where 
\begin{eqnarray}
FA=-27\,{\frac { \left( -\beta_{{0}}c+c \right)  \left( \beta_{{0}}c+c
 \right)  \left( c\sqrt {-{\beta_{{0}}}^{2}{c}^{2}+{c}^{2}}-{c}^{2}+{
\beta_{{0}}}^{2}{c}^{2} \right) \sqrt {3}{b}^{5}t}{ \left( 3\,{b}^{2}+
{r_{{0}}}^{2} \right) ^{5/2}\sqrt {-{\beta_{{0}}}^{2}{c}^{2}+{c}^{2}}r
_{{0}}c}}+\beta_{{0}}c
\nonumber \\ 
+27\,{\frac { \left( -\beta_{{0}}c+c \right) 
 \left( \beta_{{0}}c+c \right)  \left( c\sqrt {-{\beta_{{0}}}^{2}{c}^{
2}+{c}^{2}}-{c}^{2}+{\beta_{{0}}}^{2}{c}^{2} \right) \sqrt {3}{b}^{5}{
\it t0}}{ \left( 3\,{b}^{2}+{r_{{0}}}^{2} \right) ^{5/2}\sqrt {-{\beta
_{{0}}}^{2}{c}^{2}+{c}^{2}}r_{{0}}c}}
\quad ,
\end{eqnarray}
and
\begin{equation}
FB =
-243\,{\frac {{c}^{4}{b}^{5} \left( \beta_{{0}}-1 \right)  \left( 
\beta_{{0}}+1 \right)  \left( {\beta_{{0}}}^{2}+\sqrt {-{\beta_{{0}}}^
{2}+1}-1 \right) }{ \left( 3\,{b}^{2}+{r_{{0}}}^{2} \right) ^{5/2}
\sqrt {-{\beta_{{0}}}^{2}+1}r_{{0}}}}
\quad .
\end{equation}
The time dilation is therefore
\begin{equation}
\tau^* = F(t;r_0,\beta_0,t_0,b)
     -   F(t_0;r_0,\beta_0,t_0,b)
\quad .
\nonumber
\end{equation}
A measure  of the time dilation
is  
\begin{equation}
D = \frac{\tau^*}{t-t_0}
\quad ,
\nonumber
\end {equation}
with   $0 < D < 1 $.
The time dilation  is displayed as function of the time 
in Figure \ref{dgraph}
and as a function of time and scaling 
in Figure \ref{matrixgraph}.

\subsection{Radioactivity}
The  decay of a radioactive  isotope is   modeled by the
following law for particles, which are in 
the laboratory frame
\begin{equation}
N(t) = N_0 e^{-\frac{(t-t_0)}{\tau}}
\nonumber
\label{ntradioactivelab}
\quad ,
\end{equation}
where
$\tau$ is the proper lifetime,
$N_0$  is the number of nuclei at  $t=t_0$
and the  half-life is  $T_{1/2}= ln(2) \; \tau$.
In a frame that  is moving with the shell,
the decay law is
\begin{equation}
N(t) = N_0 e^{-\frac{\tau^*}{\tau}}
\nonumber
\label{ntradioactivesn}
\quad .
\end{equation}
Let us analyze the 
isotope $^{56}$Ni for which 
$\tau $ = 8.757 d  or  $T_{1/2}$ =6.07 d.
We now express the proper lifetime in yr (1yr=365.24219 d),
$\tau=0.024$
and  Table \ref{radio} reports 
the number of nuclei  that have survived at a given time. 
\begin{table}[ht!]
\caption 
{ 
Parameters of the radioactive decay 
for the isotope $^{56}$Ni
when  t=0.006535~yr
and the other parameters are the same as in  Figure \ref{rt_emd}.
}
\label{radio}
\begin{center}
\begin{tabular}{|c|c|c|}
\hline
parameter  &  no~time~dilation     & time~dilation        \\
\hline     
N          & $N=N_0\times 0.8727$  & $N=N_0\times0.9415$  \\
\hline
\end{tabular}
\end{center}
\end{table}
>From the above  table, it is evident that the 
number of nuclei that are embedded in the moving layer  
is bigger  when  the  time dilation is considered.

\section{Conclusions}

The  kinetic energy conservation for  an expansion  in the framework 
of the thin layer approximation has been extended to  SR.
We analyzed four  types of CSM and 
we derived the equation for the numerical trajectory.
A Taylor expansion for the trajectory 
has been derived in each of the cases that are modeled by  
constant, power law,  exponential
and Emden profile.
The numerical results 
were applied to the  real data of \snr1993j. 
The best results are obtained for a power law 
dependence of the CSM with $\alpha=2.15$.
The case of an expansion in a medium with 
constant density is not compatible with the data
of \snr1993j.
Some evaluations of  time dilation and of radioactivity 
in the early phase
of expansion have been done using the  Taylor expansion 
for the trajectory. 
Here we   processed  as astrophysical object only  \snr1993j; 
the  connection between SNs  and  Gamma Ray Bursts 
is demanded to a forthcoming analysis. 


\begin{thebibliography}{10}
\expandafter\ifx\csname url\endcsname\relax
  \def\url#1{{\tt #1}}\fi
\expandafter\ifx\csname urlprefix\endcsname\relax\def\urlprefix{URL }\fi
\providecommand{\eprint}[2][]{\url{#2}}

\bibitem{Wieland2016}
{Wieland} V, {Pohl} M, {Niemiec} J and et~al 2016 {Nonrelativistic
  Perpendicular Shocks Modeling Young Supernova Remnants: Nonstationary
  Dynamics and Particle Acceleration at Forward and Reverse Shocks} {\em
  \apj\/} {\bf 820}(1) 62 (\textit{Preprint} \eprint{1602.05064})

\bibitem{Bohdan2019a}
{Bohdan} A, {Niemiec} J, {Pohl} M and et~al 2019 {Kinetic Simulations of
  Nonrelativistic Perpendicular Shocks of Young Supernova Remnants. I. Electron
  Shock-surfing Acceleration} {\em \apj\/} {\bf 878}(1) 5 (\textit{Preprint}
  \eprint{1904.13153})

\bibitem{Bohdan2019b}
{Bohdan} A, {Niemiec} J, {Pohl} M and et~al 2019 {Kinetic Simulations of
  Nonrelativistic Perpendicular Shocks of Young Supernova Remnants. II.
  Influence of Shock-surfing Acceleration on Downstream Electron Spectra} {\em
  \apj\/} {\bf 885}(1) 10 (\textit{Preprint} \eprint{1909.05294})

\bibitem{Bohdan2020}
{Bohdan} A, {Pohl} M, {Niemiec} J and et~al 2020 {Kinetic Simulations of
  Nonrelativistic Perpendicular Shocks of Young Supernova Remnants. III.
  Magnetic Reconnection} {\em \apj\/} {\bf 893}(1) 6 (\textit{Preprint}
  \eprint{2003.01879})

\bibitem{Taub1948}
{Taub} A~H 1948 {Relativistic Rankine-Hugoniot Equations} {\em Physical
  Review\/} {\bf 74}, 328

\bibitem{Blandford1976}
{Blandford} R~D and {McKee} C~F 1976 {Fluid dynamics of relativistic blast
  waves} {\em Physics of Fluids\/} {\bf 19}, 1130

\bibitem{Yokosawa1984}
{Yokosawa} M 1984 {Reverse Shock Wave in Relativistic Explosions} {\em \apss\/}
  {\bf 107}(1), 109

\bibitem{Ellison1991}
{Ellison} D~C and {Reynolds} S~P 1991 {Electron acceleration in a nonlinear
  shock model with applications to supernova remnants} {\em \apj\/} {\bf 382},
  242

\bibitem{Nakar2012}
{Nakar} E and {Sari} R 2012 {Relativistic Shock Breakouts. A Variety of
  Gamma-Ray Flares: From Low-luminosity Gamma-Ray Bursts to Type Ia Supernovae}
  {\em \apj\/} {\bf 747} 88

\bibitem{Ohtani2013}
{Ohtani} Y, {Suzuki} A and {Shigeyama} T 2013 {Generation of High-energy
  Photons at Ultra-relativistic Shock Breakout in Supernovae} {\em \apj\/} {\bf
  777}(2) 113 (\textit{Preprint} \eprint{1309.1239})

\bibitem{Zhao2015}
{Zhao} X, {Wang} X, {Maeda} K and et~al 2015 {The Silicon and Calcium
  High-velocity Features in Type Ia Supernovae from Early to Maximum Phases}
  {\em \apjs\/} {\bf 220}(1) 20 (\textit{Preprint} \eprint{1508.02042})

\bibitem{Zaninetti2011a}
{Zaninetti} L 2011 {Time-dependent models for a decade of SN 1993J} {\em
  \apss\/} {\bf 333}, 99

\bibitem{Zaninetti2014f}
{Zaninetti} L 2014 { A classical and a relativistic law of motion for spherical
  supernovae } {\em \apj\/} {\bf 795}, 80

\bibitem{Marcaide2009}
{Marcaide} J~M, {Mart{\'{\i}}-Vidal} I, {Alberdi} A and {P{\'e}rez-Torres} M~A
  2009 {A decade of SN 1993J: discovery of radio wavelength effects in the
  expansion rate} {\em \aap\/} {\bf 505}, 927 (\textit{Preprint}
  \eprint{0903.3833})

\bibitem{Zaninetti2020a}
{Zaninetti} L 2020 {Energy Conservation in the Thin Layer Approximation: I. The
  Spherical Classic Case for Supernovae Remnants} {\em International Journal of
  Astronomy and Astrophysics\/} {\bf 10}(2), 71 (\textit{Preprint}
  \eprint{2004.14869})

\bibitem{Freund2008}
{Freund} J 2008 {\em {Special Relativity for Beginners: a Textbook for
  Undergraduates}\/} (Singapore: World Scientific Press)

\bibitem{Lane1870}
{Lane} H~J 1870 On the theoretical temperature of the sun, under the hypothesis
  of a gaseous mass maintaining its volume by its internal heat, and depending
  on the laws of gases as known to terrestrial experiment {\em American Journal
  of Science\/} {\bf 148}, 57

\bibitem{Emden1907}
{Emden} R 1907 {\em Gaskugeln: anwendungen der mechanischen w{a}rmetheorie auf
  kosmologische und meteorologische probleme\/} (Berlin: B. Teubner.)

\bibitem{Einstein1905}
{Einstein} A 1905 {Zur Elektrodynamik bewegter K{o}rper} {\em Annalen der
  Physik\/} {\bf 322}, 891

\end{thebibliography}
\providecommand{\newblock}{}

\end{document}